\documentclass[%
 aip,
 amsmath,amssymb,
 reprint,%
]{revtex4-1}
\usepackage[utf8]{inputenc} %
\usepackage[T1]{fontenc}    %
\usepackage[colorlinks,
    linkcolor=citelightblue,citecolor=citelightblue,urlcolor=citelightblue]{hyperref}       %
\usepackage{booktabs}       %
\usepackage{soul}
\usepackage[table,xcdraw]{xcolor}
\usepackage{graphicx}
\usepackage{threeparttable}
\usepackage{multirow}
\usepackage{amsmath}
\usepackage{bm}
\usepackage{bbm}
\usepackage{amsthm}
\usepackage{enumitem} %
\usepackage[subrefformat=parens,labelformat=parens]{subfig}

\usepackage{wrapfig}
\usepackage[ruled, vlined, linesnumbered, noend]{algorithm2e}

\definecolor{citecolor}{RGB}{34,139,34}
\definecolor{mydarkblue}{rgb}{0,0.08,1}
\definecolor{mydarkgreen}{rgb}{0.02,0.6,0.02}
\definecolor{mydarkred}{rgb}{0.8,0.02,0.02}
\definecolor{mydarkorange}{rgb}{0.40,0.2,0.02}
\definecolor{mypurple}{RGB}{111,0,255}
\definecolor{myred}{rgb}{1.0,0.0,0.0}
\definecolor{mygold}{rgb}{0.75,0.6,0.12}
\definecolor{myblue}{rgb}{0,0.2,0.8}
\definecolor{mydarkgray}{rgb}{0.,0.2,0.2}

\definecolor{lightred}{RGB}{255,235,235}
\definecolor{lightgreen}{RGB}{235,255,235}
\definecolor{lightblue}{RGB}{235,235,255}
\definecolor{citelightblue}{RGB}{49,164,222}
\definecolor{lightcyan}{RGB}{235,255,255}
\definecolor{lightmagenta}{RGB}{255,235,255}
\definecolor{lightyellow}{RGB}{255,255,235}

\definecolor{qxkcolor}{RGB}{215,235,255}
\definecolor{softmaxcolor}{RGB}{230,235,255}
\definecolor{probxvcolor}{RGB}{255,255,235}

\definecolor{topkcolor}{RGB}{255,235,235}
\definecolor{zecolor}{RGB}{255,255,235}
\definecolor{dynacolor}{RGB}{235,255,255}

\definecolor{reviewcolor}{RGB}{0,0,200}

\newcommand{\floor}[1]{\lfloor #1 \rfloor}
\newcommand{\ceil}[1]{\lceil #1 \rceil}

\newcommand{\calN}{\mathcal{N}}
\newcommand{\calL}{\mathcal{L}}

\DeclareMathOperator*{\argmin}{argmin}

\theoremstyle{plain}

\theoremstyle{definition}

\newcommand{\lemon}{\texttt{M\textsuperscript{3}ICRO}\xspace}

\usepackage{graphicx}%
\usepackage{dcolumn}%
\usepackage{bm}%

\usepackage[utf8]{inputenc}
\usepackage[T1]{fontenc}
\usepackage{mathptmx}
\usepackage{etoolbox}

\makeatletter
\def\@email#1#2{%
 \endgroup
 \patchcmd{\titleblock@produce}
  {\frontmatter@RRAPformat}
  {\frontmatter@RRAPformat{\produce@RRAP{*#1\href{mailto:#2}{#2}}}\frontmatter@RRAPformat}
  {}{}
}%
\makeatother
\begin{document}

\title[]{Differentiable Hardware-in-the-Loop Optimization\\ Enabling Versatile Photonic Neuromorphic Engines}
\title[]{Machine Learning-Enabled Photonic Subspace Neural Network based on Programmable Multi-Operand Multimode Interference}
\title[]{M\textsuperscript{3}ICRO: \underline{M}achine Learning-Enabled \underline{C}ompact Photonic Tensor Core based on P\underline{R}ogrammable \underline{M}ulti-\underline{O}perand \underline{M}ultimode \underline{I}nterference}

\author{Jiaqi Gu}
\email{jiaqigu@asu.edu.}
\affiliation{ 
Department of Electrical and Computer Engineering, The University of Texas at Austin, Austin, TX 78712, USA%
}%
 \affiliation{School of Electrical, Computer and Energy Engineering, Arizona State University, Tempe, AZ 85287, USA}%
\author{Hanqing Zhu}%
\affiliation{ 
Department of Electrical and Computer Engineering, The University of Texas at Austin, Austin, TX 78712, USA%
}%
\author{Chenghao Feng}%
\affiliation{ 
Department of Electrical and Computer Engineering, The University of Texas at Austin, Austin, TX 78712, USA%
}%
\author{Zixuan Jiang}%
\affiliation{ 
Department of Electrical and Computer Engineering, The University of Texas at Austin, Austin, TX 78712, USA%
}%
\author{Ray T. Chen}%
\affiliation{ 
Department of Electrical and Computer Engineering, The University of Texas at Austin, Austin, TX 78712, USA%
}%
\author{David Z. Pan}%
\affiliation{ 
Department of Electrical and Computer Engineering, The University of Texas at Austin, Austin, TX 78712, USA%
}%

\date{\today}%

\begin{abstract}

Photonic computing shows promise for transformative advancements in machine learning (ML) acceleration, offering ultra-fast speed, massive parallelism, and high energy efficiency. 
However, current photonic tensor core (PTC) designs based on standard optical components hinder scalability and compute density due to their large spatial footprint. 
To address this, we propose an ultra-compact PTC using customized programmable multi-operand multimode interference (MOMMI) devices, named \lemon. 
The programmable MOMMI leverages the intrinsic light propagation principle, providing a single-device programmable matrix unit beyond the conventional computing paradigm of one multiply-accumulate (MAC) operation per device. 
To overcome the optimization difficulty of customized devices that often requires time-consuming simulation, we apply ML for optics to predict the device behavior and enable a differentiable optimization flow. 
We thoroughly investigate the reconfigurability and matrix expressivity of our customized PTC, and introduce a novel block unfolding method to fully exploit the computing capabilities of a complex-valued PTC for near-universal real-valued linear transformations. 
Extensive evaluations demonstrate that \lemon achieves a 3.5-8.9$\times$ smaller footprint, 1.6-4.4$\times$ higher speed, 9.9-38.5$\times$ higher compute density, 3.7-12$\times$ higher system throughput, and superior noise robustness compared to state-of-the-art coherent PTC designs. It also outperforms electronic digital A100 GPU by 34.8-403$\times$ higher throughput while maintaining close-to-digital task accuracy across various ML benchmarks.
\end{abstract}

\maketitle

\section{Introduction}
\label{sec:Introduction}
\begin{figure}
    \centering
    \includegraphics[width=\columnwidth]{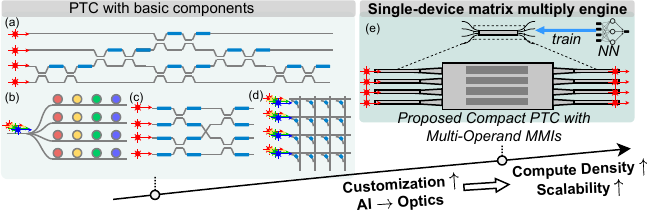}
    \caption{Overview of photonic tensor core designs with increasing compute density.
    PTCs with standard devices: (a) MZI array~\cite{NP_NATURE2017_Shen}, (b) MRR weight bank~\cite{NP_SciRep2017_Tait}, (c) Butterfly-style PTC~\cite{NP_ASPDAC2020_Gu}, and (d) PCM crossbar~\cite{NP_Nature2021_Feldmann}.
    (e) Our proposed \lemon PTC with customized MMI devices and trained with a machine learning-based approach. }
    \label{fig:Overview}
    \vspace{-10pt}
\end{figure}

\vspace{-10pt}
Photonic computing has emerged as a promising technology for high-performance and energy-efficient computing, particularly in computation-intensive artificial intelligence (AI) applications~\cite{NP_NATURE2017_Shen,NP_PIEEE2020_Cheng, NP_Nature2020_Wetzstein, NP_NaturePhotonics2021_Shastri,NP_Nature2021_Xu, NP_Nature2021_Feldmann,NP_NatureElectronics2021_Huang,NP_NatureComm2023_Bai, NP_NatureComm2023_Fu,NP_NaturePhotonics2021_Zhou}. 
Photonic tensor cores (PTCs) have been developed using standard optical components to enable matrix multiplication in the analog domain at the speed of light, including free-space diffractive designs~\cite{NP_NaturePhotonics2021_Zhou} and integrated photonic circuit-based designs~\cite{NP_NATURE2017_Shen,NP_Nature2021_Xu,NP_NatureElectronics2021_Huang,NP_NatureComm2023_Bai}. 
However, concerns regarding area efficiency and scalability arise due to the large number of bulky components used in existing PTC designs, shown in Fig.~\ref{fig:Overview}(a-d).
Based on matrix decomposition, general matrix multiplication (GEMM), i.e., universal linear operations, can be mapped to cascaded Mach-Zehnder interferometer (MZI) arrays~\cite{NP_NATURE2017_Shen}.
The large number of bulky MZIs used in the tensor core raises concerns about area efficiency and scalability.
Efforts have been made to reduce the circuit footprint through approaches such as butterfly-style photonic mesh~\cite{NP_ASPDAC2020_Gu, NP_TCAD2022_Gu, NP_ACSPhotonics2022_Feng} with logarithmic network depth, automatically searched circuit topologies~\cite{NP_DAC2022_Gu}, and low-rank MZI arrays~\cite{NP_APLPhotonics2021_Xiao}. 
There are also integrated diffractive optical neural networks (DONNs) that leverages on-chip diffractive components for high-parallelism computing~\cite{NP_NatureComm2022_Zhu, NP_NatureComm2023_Fu, NP_NatureComm2022_Wang}.
Besides, incoherent PTCs based on micro-ring resonator (MRR) weight banks~\cite{NP_SciRep2017_Tait, NP_DATE2019_Liu, NP_ISVLSI2022_Sunny,NP_DAC2021_Febin}, phase-change material (PCM) crossbar arrays~\cite{NP_APR2020_Miscuglio,NP_Nature2021_Feldmann}, and frequency micro-comb~\cite{} have been proposed for compact GEMM using multiple wavelengths.
However, the above works are based on standard components designed for optical communications.
Their compute density is still limited by approximately one multiply-accumulate (MAC) per device, which intrinsically limits their scalability and efficiency.

To address the limitations of current PTC designs and enhance area efficiency, customized photonic devices tailored for optical computing have attracted attention. 
Multi-operand (MO) photonic devices have been explored to increase compute density.
A compact photonic neuron based on multi-operand rings (MORRs)~\cite{NP_DATE2021_Gu2, NP_TCAD2022_Gu} was proposed to squeeze vector dot-product and Lorentzian nonlinear transmission into a single MORR by setting multiple controllers inside the ring, i.e., $y=f(\sum_i\phi_i(w_ix_i^2))$.
Compared to single-operand MRR weight banks~\cite{NP_DATE2019_Liu, NP_SciRep2017_Tait}, MORR arrays can significantly reduce ring resonator and wavelength usage.
In the multi-operand device family, another member, multi-operand MZI (MOMZI)~\cite{NP_arXiv2023_Feng}, was recently presented to partition the phase shifter in the MZI for vector dot-product with a sinusoidal nonlinear transmission, i.e., $y=\cos(\sum_i\phi(w_ix_i))$.
By squeezing vector/tensor operations into a single device, multi-operand devices represent a new design paradigm to scale up the compute density of optical computing.
However, for previous multi-operand devices, inputs and weights are encoded as the electrical control signals and controller tuning coefficients, respectively.
Hence, they face challenges in limited weight reconfigurability and trainability difficulties associated with nonlinear transmission.

To achieve a breakthrough in the area efficiency compared to coherent PTCs based on basic devices while overcoming the limitation of existing multi-operand PTCs, we propose a novel coherent multi-path PTC design \lemon based on customized programmable multi-operand multimode interference (MOMMI) devices, shown in Fig.~\ref{fig:Overview}(e).
By leveraging the principles of light propagation and interference, combined with fine-grained refractive index tuning within the multimode waveguide, MOMMIs enable the realization of ultra-compact programmable analog matrix multiplication cores.
Our proposed PTC, equipped with a machine learning-enabled training flow and a block unfolding method, facilitates efficient and differentiable training of complex-valued coherent PTCs based on customized devices and supports real-valued linear operations.

The contributions of \lemon are summarized as follows,
\begin{itemize}
    \item \textbf{Closing the Loop of Photonics for AI and AI for Photonics} -- We propose the first ML-enabled programmable photonic tensor core (PTC) based on customized optical devices.
    \item \textbf{Ultra-Compact Single-Device Optical Matrix Unit} -- We introduce an ultra-compact photonic tensor core based on customized programmable MOMMIs, a single-device matrix unit beyond the conventional paradigm of one MAC/device, significantly improving compute density and area efficiency.
    \item \textbf{Superior Expressivity and Footprint Efficiency} -- We enhance the expressivity of MOMMIs by developing a multi-path PTC architecture called \lemon, offering superior matrix representability and improvements in footprint efficiency over previous coherent PTCs.
    \item \textbf{ML-Assisted PTC Training Method} -- We propose a novel ML-assisted training method that estimates device gradients and enables differentiable optimization of MOMMIs, eliminating the need for time-consuming simulations and accelerating the training process.
    \item \textbf{Efficient Complex PTCs with Block Unfolding} -- We introduce a novel block unfolding technique, achieving efficient, full-range, real-to-real linear transformations with 4x higher efficiency than previous differential photodetection approaches.
    \item \textbf{Significant Performance Advantages} -- Extensive evaluations show that our customized \lemon PTC demonstrates near-universal matrix expressivity, close-to-digital accuracy on various ML tasks with 3.5-8.9$\times$ smaller footprint, 1.6-4.4$\times$ higher speed (TOPS), 9.9-38.5$\times$ higher compute density (TOPS/mm\textsuperscript{2}), 3.7-12$\times$ higher system throughput (FPS), and superior noise robustness than prior state-of-the-art (SoTA) coherent PTCs. 
    It also outperforms electronic digital Nvidia A100 GPU by 34.8-403$\times$ higher throughput. These results highlight the potential of device customization for advancing scalable photonic ML computing.
\end{itemize}

\vspace{-15pt}
\section{Proposed MOMMI-based PTC \lemon}
\label{sec:MMIPTC}
\vspace{-10pt}
We introduce a compact photonic tensor core (PTC) design \lemon based on customized programmable multi-operand MMI devices.
We design a programmable MOMMI and investigate its matrix expressivity.
Based on it, we construct the multi-path PTC \lemon with a compact footprint and near-universal matrix representability. 
We introduce an efficient ML-based training method for customized photonic devices. 
Additionally, we present a novel block unfolding method to overcome optimization challenges in complex coherent PTCs.

\vspace{-15pt}
\subsection{Initial State Design of General MMI Device}
\label{sec:MMIDesign}
\vspace{-10pt}
We start our PTC design from an initial MMI structure with a compact footprint, low insertion loss, and near-uniform power splitting ratios.
This requires us to carefully determine the width and length of the MMI.
Consider a 2-dimensional (2-D) horizontal plane of an MMI, we denote its length as $L$, width as $W_{MMI}$, effective refractive index of the multimode region as $n_{eff}$, and index of the cladding as $n_{c}$.
We define $L_{\pi}\approx\frac{4n_{eff}W_{e0}^2}{3\lambda_0}$, where $W_{e0}$ is the effective width of the 0-th mode.
Based on the dispersion equation~\cite{NP_JLT1995_Soldano}, we obtain the propagation constant spacing between the 0-th and $v$-th mode as $\beta_0-\beta_{\nu}\approx \frac{\nu(\nu+2)\pi}{3L_{\pi}}$.
The field profile $\Psi{(y,z)}$ at the output ports can be written as the superposition of all guided modes at $z=L$, $\Psi(y,L)=\sum_{\nu=0}^{m-1}c_{\nu}\psi_{\nu}(y)\exp{\Big[jL\frac{\nu(\nu+2)\pi}{3L_{\pi}}\Big]}$.
The output field should be a multiple self-imaging of the input field $\Psi(y,0)$, which holds at the condition of
\begin{equation}
    \label{eq:SelfImaging}
    \small
    \exp\Big[jL\frac{\nu(\nu+2)\pi}{3L_{\pi}}\Big]=1 \text{ or } (-1)^{\nu},~~ L=p(3L_{\pi}/N),~~ p\in\mathbb{Z}.
\end{equation}
\begin{figure}[htp]
    \centering
    \includegraphics[width=0.9\columnwidth]{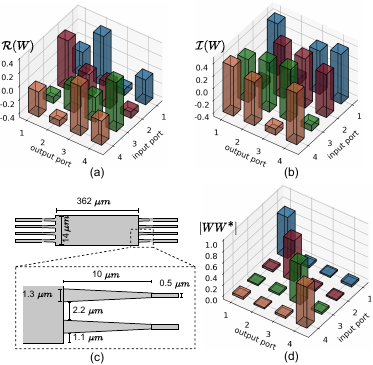}
    \caption{~(a) Real and (b) imaginary parts of the transfer matrix of the optimized 4$\times$4 MMI.
    (c) Detailed sizes of the MMI.
    (d) The transfer matrix of the optimized MMI is close to a unitary matrix.}
    \label{fig:MMI4x4_Matrix}
\end{figure}

To obtain an MMI with the shortest length, we set $p=1$ and $L=3L_{\pi}/N$, which corresponds to the first $N$-fold self-imaging.
Based on this initialization, we simulate the figure of merit (FoM) of the MMI, defined as the product of the insertion loss and imbalance of power splitting, while performing hyperparameter search on $L$ and $W_{MMI}$ to optimize the FoM.
Ideally, the transfer function of a general $k \times k$ MMI corresponds to a symmetric unitary matrix, given its geometric symmetry and energy conservation.
For example, after device optimization, the spatial dimensions and transfer matrix of our optimized 4$\times$4 MMI are shown in Fig.~\ref{fig:MMI4x4_Matrix}.
We observe a nearly symmetric unitary transfer matrix and a near-uniform power splitting ratio, which is a good initial state of the MMI.

\begin{figure}
    \centering
    \includegraphics[width=0.85\columnwidth]{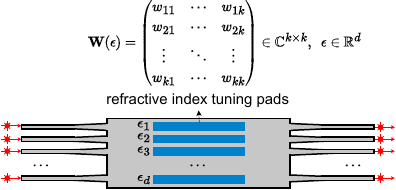}
    \caption{A $d$-op $k\times k$ programmable multi-operand MMI.}
    \label{fig:ProgrammableMMI}
\end{figure}

\vspace{-15pt}
\subsection{\lemon: Programmable MOMMI-based PTC}
\vspace{-10pt}
Now we discuss how to make an MMI reprogrammable, and then we will introduce how to construct our \lemon tensor core using this customized device. 

\begin{figure}
    \centering
    \includegraphics[width=\columnwidth]{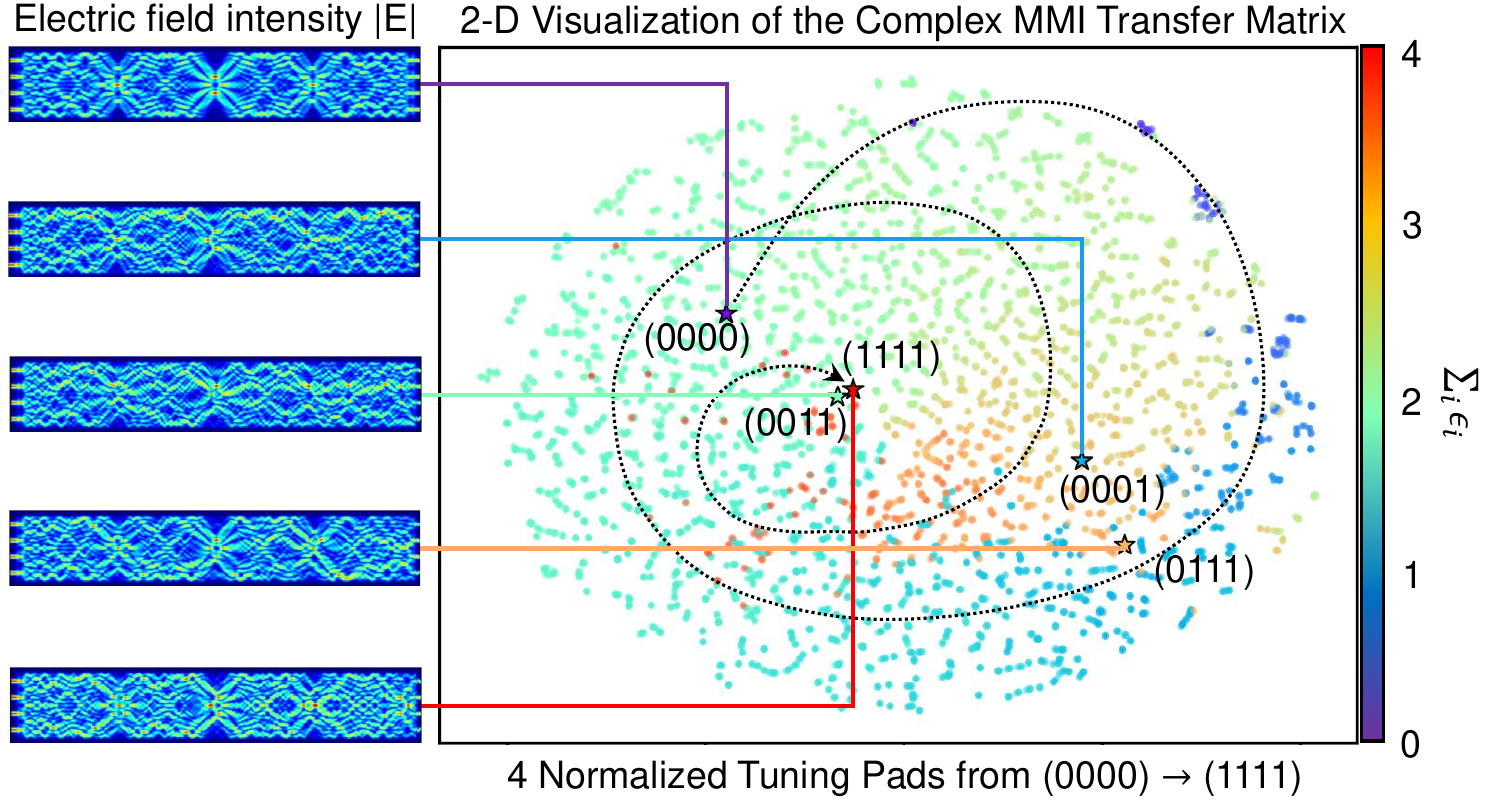}
    \caption{Visualization of the complex transfer matrix $W{(\epsilon)}\in\mathbb{C}^{4\times 4}$ of a 4-op 4$\times$4 MOMMI in the projected 2-D space using t-SNE.
    Each pad is discretized to 8 uniform levels (3-bit) and normalized to [0,1] by the maximum index change (0.03).
    Matrices are colored based on $\sum_i\epsilon_i$.}
    \label{fig:MMI_MatrixSpace}
\end{figure}

\noindent\textbf{Programmable MOMMI}.~By changing the refractive index inside the multimode waveguide region, we can program the transfer matrix of the MMI.
As shown in Fig.~\ref{fig:ProgrammableMMI}, we introduce a customized multi-operand MMI (MOMMI) by placing $d$ tunable regions within an MMI to change their local refractive indices ($\epsilon_1, \cdots, \epsilon_d$).
In this way, we can perform fine-grained manipulation of the device transmission.
Discussion on the practicality of the device implementation is in Section~\ref{sec:DeviceImplementation}.
Note that the complex-valued transfer matrix $W(\epsilon)$ of a $d$-op $k\times k$ MOMMI is reparametrized by $d$ refractive indices, leading to a reduced degree of freedom with only $d$ real latent variables.
Therefore the representable matrices are restricted to a subspace of arbitrary complex matrices.
We sweep the refractive indices for each tuning pad and visualize the simulated transfer matrices of a 4-op 3-bit 4$\times 4$ MOMMI in Fig.~\ref{fig:MMI_MatrixSpace}.
A clear spiral-like matrix distribution in the parameter space can be observed as we gradually increase the normalized indices from (0,0,0,0) to (1,1,1,1) with 3-bit resolution on each pad, which represents the implementable matrix subspace.

A single MOMMI itself is an ultra-compact matrix unit.
However, 
with a reduced number of parameters ($d<2k^2$), it shows limited expressivity and lacks flexible controllability over matrix norm and signs, evidenced in Fig.~\ref{fig:MMI_MatrixSpace}.
Therefore, we need to enhance its expressivity with a specialized tensor core design.

\begin{figure}
    \centering
    \includegraphics[width=\columnwidth]{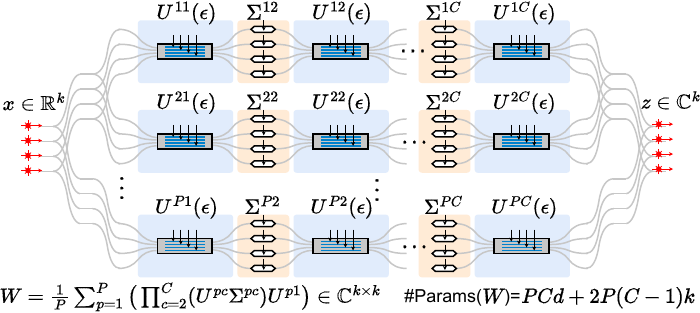}
    \caption{The proposed MOMMI-based photonic tensor core \lemon with $P$ parallel paths and $C$ cascaded components.
    }
    \label{fig:MOMOMI_PTC}
\end{figure}

\begin{figure*}
    \centering
    \subfloat[]{\includegraphics[width=0.52\textwidth]{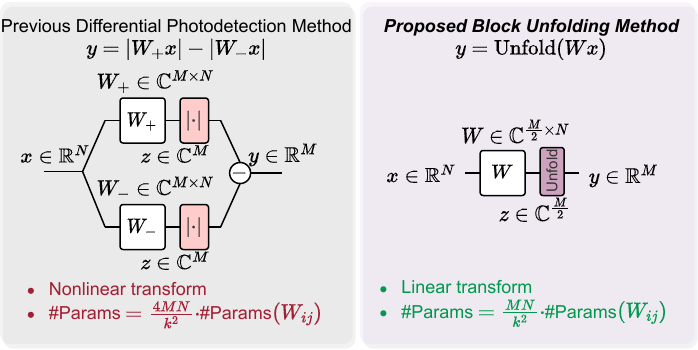}
    \label{fig:ComparePTC}
    }
    \subfloat[]{\includegraphics[width=0.39\textwidth]{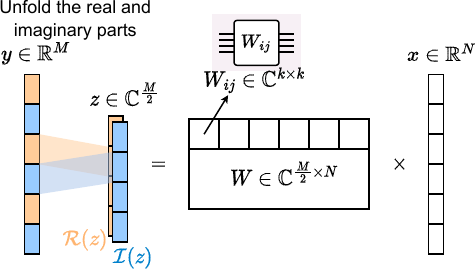}
    \label{fig:BlockUnfold}
    }
    \vspace{-5pt}
    \caption{(a) Compared to previous complex-valued photonic tensor core designs with differential photodetection, our proposed block unfolding method supports pure linear transform with 4 times fewer parameters.
    (b) Illustration of block unfolding that interleaves the output vector's real part and imaginary part block-wise.
    }
    \label{fig:Unfolding}
\end{figure*}

\noindent\textbf{Multi-Path PTC \lemon}.~
To enhance the matrix expressivity, we introduce a multi-path PTC \lemon in Fig.~\ref{fig:MOMOMI_PTC}, constructed by cascading $C$ blocks of interleaved MOMMIs and modulators with $P$ parallel paths.
Each MOMMI serves as an all-to-all channel mixer to create dense signal interactions. 
The internal diagonal complex matrix $\Sigma\in\mathbb{C}^{k}$ handles row-column scaling to modulate the matrix norm and signs.
The formulation of the multi-path PTC \lemon is as follows,
\begin{equation}
    \label{eq:PTCFormulation}
    W=\frac{1}{P}\sum_{p=1}^{P}\big(\prod_{c=2}^{C}(U^{pc}\Sigma^{pc})U^{p1}\big)\in\mathbb{C}^{k \times k}.
\end{equation}
For a $k \times k$ multi-path PTC, instead of having $2k^2$ real parameters in a general complex matrix, it has a reduced number of latent real variables, i.e., $PCd+2P(C-1)k$.
As the architecture design variables of \lemon, $P$ and $C$ can be adjusted to trade off hardware efficiency and matrix expressivity.
For example, If ($k$,$d$,$P$,$C$)=(4,4,2,2), it has exactly the same parameter count as a general complex matrix unit, i.e., 32.

\vspace{-15pt}
\subsection{Efficient Complex Tensor Core via Block Unfolding}
\label{sec:BlockUnfolding}
\vspace{-10pt}
A complex matrix unit seems to have a higher expressivity than a real counterpart since it doubles the parameter count.
However, it is often not true when applied to neural networks that require real-valued operations, e.g., activation functions, normalization, pooling, and loss functions.
Therefore, to fit into the widely used real-valued DNN paradigm, we need to construct a photonic tensor core that supports \emph{full-range real-valued inputs/outputs}.
Previous methods either (1) enforce a real transfer matrix that wastes the multiplication of the imaginary part, e.g., MZI arrays~\cite{NP_NATURE2017_Shen,NP_Arxiv2022_Demirkiran, NP_ASPDAC2019_Zhao, NP_ASPDAC2020_Gu, NP_JSTQE2023_Xiao}, or (2) remove the phase information by extracting the light intensity through photodetection, which only supports non-negative output~\cite{NP_ACSPhotonics2022_Feng, NP_NatureComm2022_Zhu, NP_NatureComm2022_Wang}.
For case (2), differential photodetection is widely used to create full-range output vectors, i.e., $y=|W_+x|-|W_-x|$, shown in Fig.~\ref{fig:ComparePTC}.
However, this method introduces undesired nonlinearity, which breaks the linear property and leads to optimization difficulty.
Moreover, such a method is not efficient as it uses two $k\times k$ complex matrix units while the \emph{effective computing} is one $k \times k$ real matrix-vector multiplication. 

To solve those problems caused by complex-valued tensor cores, we propose a block unfolding method to enable \emph{efficient, full-range, real-to-real linear transformation}.
Figure~\ref{fig:BlockUnfold} illustrates the principle of block unfolding.
For an $N$-input $M$-output real linear layer, we first construct a $\frac{M}{2} \times N$ complex matrix and partition it into a series of $k\times k$ blocks. 
Each complex submatrix $W_{ij}\in\mathbb{C}^{k\times k}$ is implemented by a $k\times k$ complex PTC.
The real and imaginary part of the output vector $z\in\mathbb{C}^{\frac{M}{2}}$ is unfolded blockwise,
\begin{equation}
    \small
    \label{eq:Unfolding}
    y=\texttt{Unfold}(z)=\big[R(z_1); I(z_1);\cdots;R(z_{N/k}); I(z_{N/k})\big]^T\in\mathbb{R}^M.
\end{equation}
Note that unfolding the output vector is equivalent to unfolding the complex weight matrix $W\in\mathbb{C}^{\frac{M}{2}\times N}$ to a 2$\times$ larger real-valued matrix $\widetilde{W}\in\mathbb{R}^{M\times N}$.
With this method, we fully leverage the actual computing capability of the tensor core with only $MN/k\cdot\text{\#Params}(W_{ij})$ parameters, which is twice more efficient than enforcing a real transfer matrix and 4 times more efficient than the differential photodetection method.
Note that this method is generic: any $k \times k$ complex-valued PTC, once equipped with our block unfolding, can support $(2k)\times k$ real matrix multiplication in one shot.
The coherent detection with phase and magnitude detection can be implemented by using self-analyzers~\cite{NP_Optica2022_Miller, NP_Science2023_Sai}.
\vspace{-15pt}
\subsection{Machine Learning-Enabled Differentiable Optimization}
\vspace{-10pt}
Optimizing customized photonic devices is challenging since it relies on time-consuming optical simulation involving Maxwell equations solving, eigenmode decomposition, and S-parameter extraction.
Such a complicated process is usually treated as a blackbox and cannot be embedded into the outer-loop NN training.
To enable the efficient optimization of the device variables $\epsilon$, we employ ML for photonics by introducing a differentiable photonic hardware estimator (DPE):
\begin{equation}
    \label{eq:NNModel}
    W_{\theta}=(y+y^T)/2;~~ y=f_{\theta}\Big(\cos\big(\omega Q(\epsilon)+\phi\big);Q(\epsilon)\Big),
\end{equation}
where $f_{\theta}(\cdot):\mathbb{R}^d\rightarrow\mathbb{C}^{k\times k}$ is a multi-layer perceptron, $Q(\epsilon)$ is the quantized refractive index, $\omega$ and $\phi$ are learnabled parameters in the predefined sinusoidal features.
The reparametrization on $W_{\theta}$ guarantees a symmetric transfer matrix based on prior knowledge.
\begin{figure}[htp]
    \centering
    \includegraphics[width=\columnwidth]{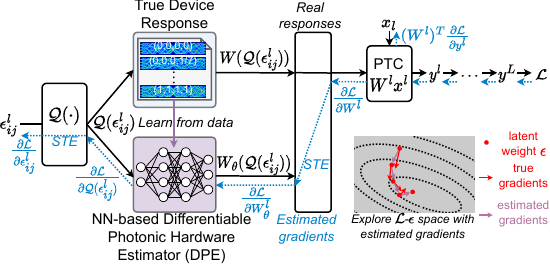}
    \caption{Proposed ML-enabled differentiable optimization flow for customized programmable MOMMI-based PTCs.}
    \label{fig:TrainMethod}
\end{figure}

\begin{figure}
    \centering
    \includegraphics[width=\columnwidth]{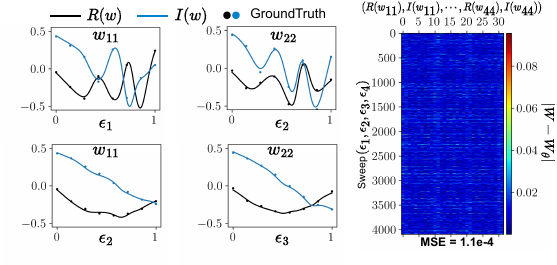}
    \caption{Visualize the prediction fidelity of our NN-based device predictor on a 4$\times$4 programmable MMI.}
    \label{fig:ModelPrediction}
\end{figure}

\begin{figure*}[]
    \centering
    \includegraphics[width=0.97\textwidth]{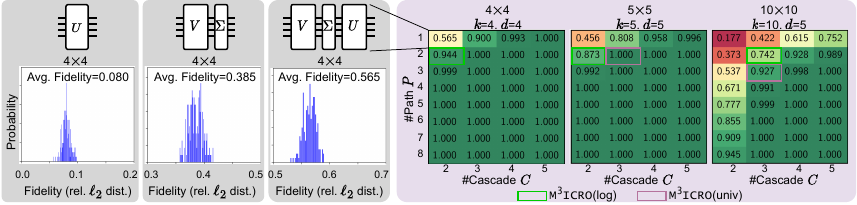}
    \caption{Numerical analysis on the matrix expressivity (fidelity) of different MOMMI-based PTC designs.
    The colormap shows how expressivity changes with different number of parallel paths $P$ and cascaded blocks $C$.}
    \label{fig:MOMMIExpressivity}
\end{figure*}

As shown in Fig.~\ref{fig:TrainMethod}, we build a differentiable training method with gradient replacement and straight-through estimator (STE) techniques.
In the forward procedure, we quantize the refractive indices to $b$-bit levels and look up the ground truth table to obtain the transfer matrix $W(\epsilon)$ for forward propagation.
During backward, we redefine the gradient calculation as follows, 
\begin{equation}
\begin{aligned}
    \frac{\partial\calL}{\partial\epsilon}=\frac{\partial\calL}{\partial W(\epsilon)}\frac{\partial W(\epsilon)}{\partial Q(\epsilon)}&\frac{\partial Q(\epsilon)}{\partial \epsilon}\approx\frac{\partial\calL}{\partial W(\epsilon)}\frac{\partial W_{\theta}}{\partial Q(\epsilon)}\\
    \frac{\partial\calL}{\partial x}&=W(\epsilon)^T\frac{\partial\calL}{\partial y}
\end{aligned},
\end{equation}
where $\frac{\partial Q(\epsilon)}{\partial \epsilon}$ is estimated as 1 using STE.
Note that only $\frac{\partial W_{\theta}}{\partial Q(\epsilon)}$ is calculated by the auto-differentiation through the NN predictor.
All other terms during forward and backward are based on $W(\epsilon)$ to eliminate gradient approximation error accumulation for higher estimation fidelity.
Figure~\ref{fig:ModelPrediction} visualizes the predicted device behavior and shows superior fidelity with 1.1e-4 mean-square error (MSE) compared to the ground-truth targets. 
Most importantly, the predictor behaves as a high-quality first-order oracle with a very smooth landscape that can provide reliable and informative first-order gradient information to guide optimization.

\vspace{-15pt}
\subsection{Expressivity of Programmable MOMMI}
\label{sec:Expressivity}
\vspace{-10pt}

\begin{figure}
    \centering
    \includegraphics[width=0.9\columnwidth]{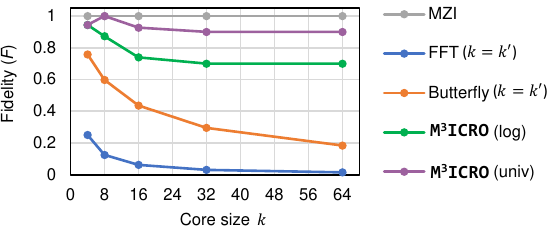}
    \caption{Compare the matrix expressivity/fidelity ($F$) on different PTC designs across various matrix sizes.
    $k'$ is the butterfly mesh size.}
    \label{fig:CompareFidelity}
\end{figure}

To evaluate the matrix expressivity of our multi-path MOMMI-based PTC \lemon, we perform numerical analysis on different PTC designs in Fig.~\ref{fig:MOMMIExpressivity}.
We randomly generate 40k real matrices from Gaussian distribution, train the differentiable surrogate model of each PTC design with block unfolding to approximate those random real matrices, and then evaluate the average relative $\ell_2$ matrix distance as the fidelity, i.e., $F=\frac{1}{N}\sum_{i=1}^{N}\|W_{\theta}(\epsilon_i)-\widetilde{W}_i\|_{\mathcal{F}}^2/\|\widetilde{W}_i\|_{\mathcal{F}}^2$. 
First of all, the diagonal matrix used for norm and phase tuning is critical to the expressivity.
From the expressivity colormap, we can conclude the following trade-off.
(1) Increasing cascading depth $C$ is more effective in boosting the expressivity, but it will significantly increase the circuit depth leading to higher delay and insertion loss.
(2) Increasing the parallel path count $P$ is not as effective in expressivity boost since it only interpolates inside the convex hull of the subspace and also introduces extra signal splitting and combining cost, but it does not increase the critical path length.
(3) With large enough photonic mesh width and depth, our \lemon can potentially realize 100\% matrix expressivity as a universal linear unit.
We also compare our \lemon with previous PTC designs in Fig.~\ref{fig:CompareFidelity} across different matrix sizes.
\lemon variants have comparable expressivity to the universal MZI array and significantly outperform previous compact PTC designs based on FFT~\cite{NP_ASPDAC2020_Gu, NP_NatureComm2022_Zhu} and trainable butterfly~\cite{NP_TCAD2020_Gu, NP_ACSPhotonics2022_Feng} topology.

\vspace{-15pt}
\subsection{Hardware Performance and Efficiency Analysis}
\label{sec:HardwareEfficiency}
\vspace{-10pt}
Section~\ref{sec:Expressivity} discussed the trade-offs between hardware cost and matrix expressivity with different depth $C$ and parallel path count $P$.
To cover two representative design points for the following discussion, we design a compact variant named \lemon(log) and a larger but more expressive variant \lemon(univ).
For \lemon(log), we prioritize area efficiency and target $\sim$70\% expressivity.
We design \lemon(log) as a dual-path PTC, i.e., $P=2$, with a logarithmic circuit depth $C=\floor{\log_2{k}}$.
For \lemon(univ), we prioritize expressivity with >90\% fidelity and design it as a near-universal PTC.
We empirically set 70\% parameter count as a target, assume $P\approx C$, $d=k$, $\alpha$=0.7 and have
\begin{equation}
    \label{eq:ParameterCount}
    \small
    \begin{aligned}
        &PCk+2P(C-1)k \approx 2\alpha k^2;\\
        &P=[{\frac{1+\sqrt{1+6\alpha k}}{3}}];C=\ceil{\frac{1+\sqrt{1+6\alpha k}}{3}}.
    \end{aligned}
\end{equation}
The following analysis mainly focuses on those two variants of \lemon architecture.

\noindent\textbf{Footprint}.~
We derive the total device footprint of a PTC as
\begin{equation}
\label{eq:Footprint}
A_{\text{total}}=A_{\text{laser}}+(k-1)A_{\text{Y}}+k A_{\text{MZM}}+A_{\text{core}}+k A_{\text{PD}},
\end{equation}
where the footprint of the computing core $A_{\text{core}}$ is derived in Appendix~\ref{sec:Appendix:Footprint_Power} and Table~\ref{tab:cost}.
$A_{\text{laser}}$, $A_{\text{Y}}$, $A_{\text{MZM}}$, and $A_{\text{PD}}$ represent the footprint of laser, Y-branch used for on-chip channel splitting, input modulators, and photodetectors.
\begin{figure*}
    \centering
    \includegraphics[width=0.86\textwidth]{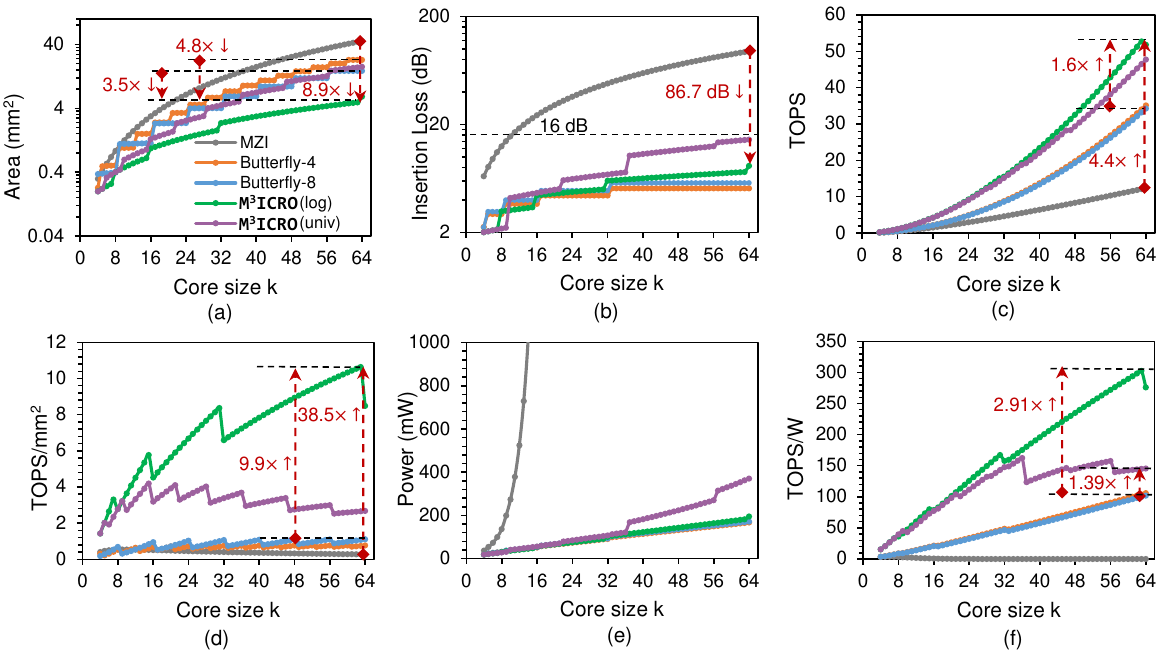}
    \vspace{-10pt}
    \caption{Compare the (a) footprint, (b) insertion loss (IL), (c) computing capacity (TOPS), (d) compute density (TOPS/mm$^2$), (e) power, and (f) energy efficiency (TOPS/W) among different PTC designs with increasing core sizes. 
    Butterfly PTCs adopt differential photodetection, and our \lemon adopts the proposed block unfolding method.
    The insertion loss excludes the input splitting and signal modulator, while others include laser, splitting, modulators, tensor cores, and photodetectors.}
    \label{fig:CompareFootprint_IL}
\end{figure*}

We plot the total footprint $A_{\text{total}}$ of different PTCs with increasing core sizes in Fig.~\ref{fig:CompareFootprint_IL}(a).
Our \lemon(log) PTC shows good footprint scalability and is 1.6$\sim$8.9$\times$ more compact than the MZI array, 1.1$\sim$4.8$\times$ smaller than FFT/Butterfly-style PTCs with a block size of 4, and 1$\sim$3.5$\times$ smaller than FFT/Butterfly PTCs with a block size of 8.
\lemon(univ) generally has a comparably compact footprint to Butterfly-8 PTC.

\noindent\textbf{Insertion Loss}.~
As an important design metric, circuit insertion loss (IL) impacts the required laser power.
High insertion loss fundamentally limits the PTC's power efficiency and scalability.
The theoretical insertion loss $IL_{\text{core}}$ in the unit of dB of different PTCs is summarized in Table~\ref{tab:cost}.
Figure~\ref{fig:CompareFootprint_IL}(b) shows the insertion loss scalability of different photonic computing cores, excluding signal splitting and input modulators.
With a 64$\times$64 core size, the MZI mesh has almost 97 dB insertion loss, while our \lemon shows less than 16 dB IL.
Such a low insertion loss of \lemon can fundamentally enable further scaling of larger core sizes with affordable laser power.

\noindent\textbf{Peak Compute Speed and Density}.~
To estimate the peak speed, we derive the PTC delay by accumulating the delay from electrical control to the final result readout~\cite{NP_ACSPhotonics2022_Feng, NN_HPCA2024_Gu_DOTA} as follows,
\begin{equation}
    \label{eq:Delay}
    \tau=\tau_{\text{EO}}+\tau_{\text{core}}+\tau_{\text{PD}}+\tau_{\text{ADC}}.
\end{equation}
We assume $\tau_{\text{EO}}$=10 ps for the electrical-to-optical (EO) conversion, 10 ps for photodetection, and 200 ps for 5 GSPS analog-to-digital conversion (ADC).
The optical path delay of the tensor core $\tau_{\text{core}}$ is derived from the total length of cascaded devices along the critical path, which is summarized in Table~\ref{tab:cost}.
The peak computing speed on a $k\times k$ matrix-vector multiplication workload is defined as $2k^2/\tau$.
Note that if our block unfolding method is applied, the peak computing speed will double, i.e., $4k^2/\tau$, as it finishes twice the computations in one shot compared to the differential detection method.
The peak computing speed (TOPS) of different PTCs with increasing core sizes is compared in Fig.~\ref{fig:CompareFootprint_IL}(c).
Our \lemon(log) has 4.4$\times$ and 1.6$\times$ faster peak computing speed than MZI arrays and Butterfly-style PTCs, respectively.
The speed can scale up when using a larger core size or wavelength-division multiplexing (WDM) for multi-wavelength parallel computing due to the broadband property of our design.

In terms of area efficiency (compute density), shown in Fig.~\ref{fig:CompareFootprint_IL}(d), since our \lemon is very compact in spatial footprint, it shows 38.5$\times$ and 9.9$\times$ higher TOPS/mm$^2$ than MZI arrays and butterfly structures, respectively.

\noindent\textbf{Power and Energy Efficiency}.~
The power of the photonic tensor core mainly consists of four parts, i.e., laser, input modulators, weight programming in the core, and photodetection,
\begin{equation}
\label{eq:power}
P_{\text{total}}=P_{\text{laser}}+P_{\text{mod}}+P_{\text{wt}}+P_{\text{PD}}.
\end{equation}
The weight programming power $P_{\text{wt}}$ is zero if using non-volatile phase shifters~\cite{NP_OE2021_Baghdadi}.
The input modulation power $P_{\text{mod}}$ and detection power $P_{\text{PD}}$ are the same for all coherent PTCs using MZMs.
Given the photodetector sensitivity $S$, ADC resolution of $b$-bit (we assume 8-bit here), and laser wall-plug efficiency $\eta$, the required wall-plug power is $P_{\text{laser}}=10^{(S+IL)/10}\times 2^b/\eta$,
where the total insertion loss $IL$ includes the loss of the computing core $IL_{\text{core}}$ (in Table~\ref{tab:cost}) and the loss of Y-branch splitting tree and input MZMs for $k$ channels, i.e.,
\begin{equation}
    \label{eq:InsertionLossTotal}
    IL=\log_{2}{k}\cdot IL_{Y}+IL_{\text{MZM}}+IL_{\text{core}}.
\end{equation}
The detailed device parameters used in the calculation are listed in Table~\ref{tab:device}.
With different core sizes $k$, we show the power consumption of different designs in Fig.~\ref{fig:CompareFootprint_IL}(e).
Butterfly-style PTCs and our \lemon have much lower insertion loss than MZI meshes, which shows considerably better power scalability to larger core sizes.
The energy efficiency is defined as the ratio of peak computing speed to power (TOPS/W).
Figure~\ref{fig:CompareFootprint_IL}(f) shows our 64$\times$64 \lemon(log) and \lemon(univ) architectures have 289.9 TOPS/W and 128.4 TOPS/W energy efficiency, outperforming Butterfly-style PTCs by 2.91$\times$ and 1.39$\times$, respectively.

\vspace{-15pt}
\section{Evaluation}
\label{sec:Results}
\vspace{-10pt}
\begin{table*}[]
\centering
\caption{Compare accuracy across different PTC designs on various models and datasets}
\label{tab:CompareAccuracy}
\resizebox{0.86\textwidth}{!}{
\begin{tabular}{l|ccccc|ccccc}
\hline
                  & MZI~\cite{NP_NATURE2017_Shen}   & FFT-4~\cite{NP_ASPDAC2020_Gu} & FFT-8~\cite{NP_ASPDAC2020_Gu} & Butterfly-4~\cite{NP_ACSPhotonics2022_Feng} & Butterfly-8~\cite{NP_ACSPhotonics2022_Feng} & \begin{tabular}[c]{@{}c@{}}\lemon   \\ (log)-4\end{tabular} & \begin{tabular}[c]{@{}c@{}}\lemon   \\ (log)-5\end{tabular} & \begin{tabular}[c]{@{}c@{}}\lemon   \\ (univ)-5\end{tabular} & \begin{tabular}[c]{@{}c@{}}\lemon   \\ (log)-10\end{tabular} & \begin{tabular}[c]{@{}c@{}}\lemon   \\ (univ)-10\end{tabular} \\\hline
ResNet20-CIFAR10  & 90.29 & 86.04 & 82.94 & 90.38       & 88.27       & 90.56        & 90.36        & \textbf{90.77}         & 90.26         & 90.10          \\
ResNet18-CIFAR100 & 73.45 & 70.88 & 67.25 & 72.63       & 72.01       & 74.18        & \textbf{74.22}        & 74.00         & 72.05         & 73.53          \\
MobileNetV3-SVHN  & 95.57 & 95.2  & 94.65 & 95.57       & 95.03       & \textbf{95.61}        & 95.38        & 95.59         & 95.19         & 95.28          \\
 \hline
\end{tabular}
}
\end{table*}

We conduct various simulation-based evaluations on our \lemon PTC designs in terms of expressivity, quantization tolerance, and noise robustness.
We mainly compare \lemon with (1) MZI array~\cite{NP_NATURE2017_Shen}, (2) FFT-based PTC with fixed optical Fourier transform modules~\cite{NP_ASPDAC2020_Gu, NP_NatureComm2022_Zhu}, and (3) Butterfly-style PTC with trainable butterfly transforms~\cite{NP_TCAD2022_Gu, NP_ACSPhotonics2022_Feng}.
Note that we do not compare with other multi-operand tensor cores since they are incoherent architectures with nonlinear transmissions and limited training scalability, especially on large NN models.
We also show the effectiveness of our block unfolding method.

\vspace{-15pt}
\subsection{Training Setups}
\label{sec:Setup}
\vspace{-10pt}
We train optical neural network models based on the open-source library \href{https://github.com/JeremieMelo/pytorch-onn}{TorchONN} and adopt the same settings for all PTC designs.
We first train a software digital NN model and use it as a teacher model $T$.
Its optical analog version is called the student model $S$.
As an initialization, we map the teacher's weight matrices $W_{ij}^T$ blockwise to the student counterpart $W_{ij}(\epsilon,\Sigma)$ by solving the optimization problem $\epsilon^{init},\Sigma^{init}=\argmin_{\epsilon,\Sigma}\sum_{i,j}\|W_{ij}^T-W_{ij}(\epsilon,\Sigma)\|_2^2$.
After mapping, we fine-tune the student with knowledge distillation, $\min_{\epsilon,\Sigma}\calL_{CE}(y^S,\hat{y})+\eta\beta^2 D_{KL}\big(\frac{y^S}{\beta},\frac{y^T}{\beta}\big)$,
where $\calL_{CE}$ is the cross-entropy loss between the student predictions and the labels, $D_{KL}$ is the KL divergence between student and teacher predictions, $\beta$ is the temperature ($\beta$=2), and $\eta$ is set to 0.1 to balance two loss functions.
During the 3000-step mapping stage, we use Adam optimizer with an initial learning rate of 1e-2 for $\Sigma$ and 1e-3 for $\epsilon$.
Cosine learning rate decay is adopted.
The fine-tuning stage learning rate is set to 3e-4 for $\Sigma$ and 4e-4 for $\epsilon$.
The NN device predictor is a 6-layer MLP: ($2k$)-(256)$_{\times 3}$-(128)$_{\times 2}$-($2k^2$).
\vspace{-15pt}
\subsection{Accuracy Evaluation}
\vspace{-10pt}
Table~\ref{tab:CompareAccuracy} shows a comprehensive comparison among different PTC designs on three different NN models and image classification datasets.
The universal MZI array represents the ideal software NN accuracy.
We observe unsatisfying FFT-based PTC due to its fixed Fourier transform and limited matrix expressivity.
The butterfly PTC has trainable phases in the butterfly transform, showing enhanced accuracy on different tasks compared to the FFT designs.
Across different MOMMI sizes, our \lemon(log) and \lemon(univ) series outperform the compact butterfly designs on all benchmarks.
Our specially designed universal \lemon variants show the best accuracy.
Even with 10$\times$10 MOMMIs, the universal variant maintains >0.9 matrix expressivity with <0.5\% accuracy degradation compared to the ideal digital software model.

\vspace{-15pt}
\subsection{Quantization Tolerance Evaluation}
\vspace{-10pt}
\begin{figure}
    \centering
    \includegraphics[width=\columnwidth]{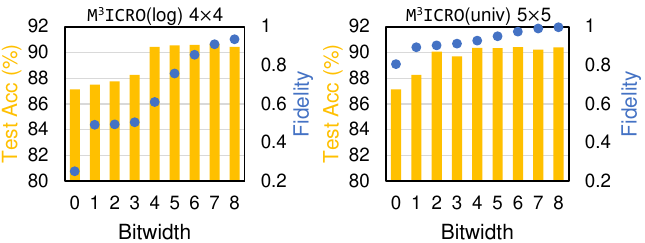}
    \caption{Matrix expressivity and test accuracy under $\epsilon$ quantization on ResNet-20 CIFAR-10 for 4$\times$4 \lemon(log) and 5$\times$5 \lemon(univ).
    0-bit means the MOMMI is fixed to its initial state.
    Activations are quantized to 8-bit.}
    \vspace{-5pt}
    \label{fig:QuantizedFidelityAccuracy}
\end{figure}

In practice, the index tuning precision inside the MOMMI is quantized for control efficiency consideration.
In Fig.~\ref{fig:QuantizedFidelityAccuracy}, we illustrate the impact of $\epsilon$ bitwidth on the PTC matrix expressivity and the corresponding accuracy on the ResNet-20 CIFAR-10 benchmark.
To stablize the optimization of discrete device control variables $\epsilon$, we set the following initial learning rate to $\min(\alpha_0, \alpha_0\times2^{b-2}), \alpha_0=5e-6$.
For \lemon(log)-4, the expressivity drops with fewer bitwidth while the task accuracy can maintain <1\% drop with above 4-bit resolution, which is suitable for efficient device control.
For \lemon(univ)-5, the fidelity maintains >0.8 even with a fixed MOMMI (0-bit), and the accuracy can almost maintain the same value with more than 2-bit.
Binary and fixed MOMMIs suffer from overly limited matrix expressivity, which further necessitates and proves the superiority of the programmability of our MOMMI device over previous passive/fixed designs~\cite{NP_NatureComm2022_Zhu, NP_ASPDAC2020_Gu}.
In practical settings, 4-bit to 8-bit resolutions are considered efficient and practical settings for most analog ML accelerators. 
Overall, our MOMMI-based PTC \lemon shows great task accuracy and quantization tolerance with 4 to 8-bit device controls.

\vspace{-15pt}
\subsection{Device Noise Robustness Evaluation}
\label{sec:NoiseRobustness}
\vspace{-10pt}

\begin{figure}
    \centering
    \includegraphics[width=\columnwidth]{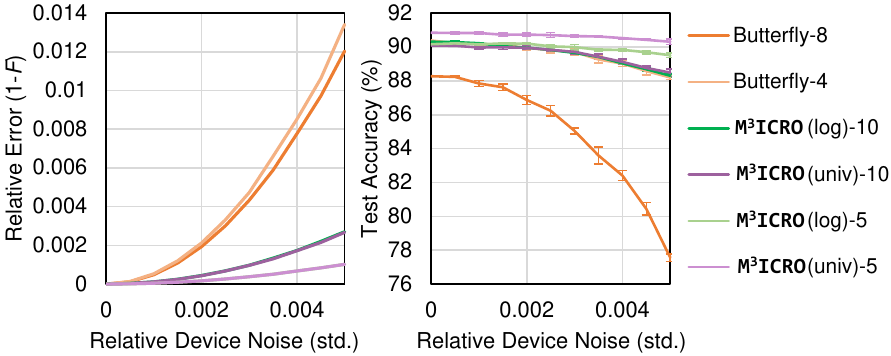}
    \caption{~Noise robustness evaluation for different tensor core designs on ResNet-20 CIFAR10 with various device noise intensity.
    All tensor cores adopt the proposed block unfolding method.
    Error bars show the accuracy standard deviation.
    The proposed \lemon-series shows superior robustness compared to previous SoTA butterfly-style PTCs.
}
    \label{fig:CompareRobustness}
\end{figure}

We evaluate the noise tolerance of our proposed \lemon PTC design against random index perturbation from nonideal control signals or environmental variations.
We mainly compare with butterfly PTC since it has the SoTA noise tolerance due to its logarithmic network depth~\cite{NP_ACSPhotonics2022_Feng, NP_TCAD2022_Gu}.
In Fig.~\ref{fig:CompareRobustness}, we first compare the relative matrix $\ell_2$ error, i.e., $1-F$, caused by various device noise intensities.
The noise is sampled from $\Delta \epsilon\sim\calN(0, \sigma^2)$ for the index of \lemon with the maximum tuning range of 1 and $\Delta\phi\sim \calN(0, (2\pi\sigma)^2)$ for the phases in the butterfly PTC with the maximum tuning range of $2\pi$~\cite{NP_ICCAD2020_Zhu}.
We observe significantly lower sensitivity of \lemon compared to butterfly designs.
We further evaluate the accuracy degradation on ResNet-20 CIFAR-10.
All \lemon variants outperform the butterfly designs with better noise tolerance.

\vspace{-15pt}
\subsection{Ablation Study on Block Unfolding}
\label{sec:CompareBlockUnfold}
\vspace{-10pt}
\begin{table}[]
\caption{Evaluate the effectiveness of our proposed block unfolding (\emph{unfold}) and previous differential photodetection (\emph{diff}) method.
\emph{div} means divergence due to the instability caused by the nonlinear absolute operations.}
\label{tab:CompareUnfold}
\resizebox{\columnwidth}{!}{
\begin{tabular}{lc|cc|cc|cc}
\hline
                                                        &          & \multicolumn{2}{c|}{FFT-4~\cite{NP_ASPDAC2020_Gu}} & \multicolumn{2}{c|}{Butterfly-4~\cite{NP_ACSPhotonics2022_Feng}} & \multicolumn{2}{c}{\lemon(log)-4} \\
                                                         &          & Diff    & \cellcolor[HTML]{EFEFEF}Unfold  & Diff    & \cellcolor[HTML]{EFEFEF}Unfold   & Diff    & \cellcolor[HTML]{EFEFEF}Unfold  \\ \hline
\multicolumn{1}{c|}{ResNet-20}                                    & Acc      & 85.17   & \cellcolor[HTML]{EFEFEF}86.04   & 86.86   & \cellcolor[HTML]{EFEFEF}90.38    & 88.44   & \cellcolor[HTML]{EFEFEF}90.56   \\
\multicolumn{1}{c|}{CIFAR100}  & \#Params & 0.27 M  & \cellcolor[HTML]{EFEFEF}67.5 K  & 1.09 M  & \cellcolor[HTML]{EFEFEF}0.27 M   & 1.09 M  & \cellcolor[HTML]{EFEFEF}0.27 M  \\ \hline
\multicolumn{1}{c|}{ResNet-18}                                    & Acc      & 67.2    & \cellcolor[HTML]{EFEFEF}70.88   & 68.77   & \cellcolor[HTML]{EFEFEF}72.63    & 70.57   & \cellcolor[HTML]{EFEFEF}74.18   \\
\multicolumn{1}{c|}{CIFAR100} & \#Params & 11.22 M & \cellcolor[HTML]{EFEFEF}2.81 M  & 44.85 M & \cellcolor[HTML]{EFEFEF}11.22 M & 44.85 M & \cellcolor[HTML]{EFEFEF}11.22 M \\ \hline
\multicolumn{1}{c|}{MobileNetV3}                                    & Acc      & div     & \cellcolor[HTML]{EFEFEF}95.2    & div     & \cellcolor[HTML]{EFEFEF}95.57    & div     & \cellcolor[HTML]{EFEFEF}95.61   \\
\multicolumn{1}{c|}{SVHN}  & \#Params & 1.53 M  & \cellcolor[HTML]{EFEFEF}0.396 M & 6.08 M  & \cellcolor[HTML]{EFEFEF}1.54 M   & 6.08 M  & \cellcolor[HTML]{EFEFEF}1.54 M  \\
 \hline
\end{tabular}
}
\end{table}

Table~\ref{tab:CompareUnfold} compares PTCs with differential photodetection and block unfolding on different benchmarks.
Differential photodetection consumes 4 times the parameters and hardware cost to perform a nonlinear real-to-real transformation with a balanced output range.
It cast significant optimization difficulty, leading to severe accuracy drop or even divergence on MobileNetV3.
In contrast, our block unfolding achieves close-to-digital accuracy because it enables a real-to-real full-range linear transform, which is compatible with direct weight matrix mapping without optimization instability issues.

\vspace{-15pt}
\subsection{Advance Compute Density vs. Efficiency Pareto Frontier}
\vspace{-10pt}
\begin{figure}
    \centering
    \includegraphics[width=0.85\columnwidth]{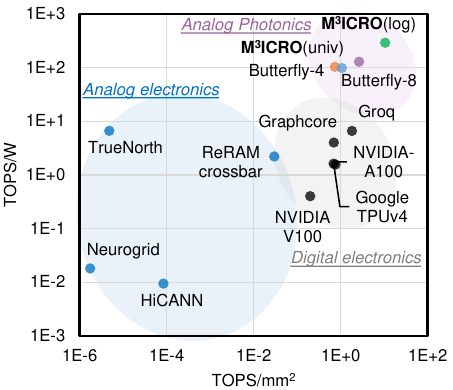}
    \caption{Compute density vs. energy efficiency Pareto frontier of different hardware technologies, including analog electronics, digital electronics, and analog photonics (64$\times$64 cores).
    Our \lemon achieves the best Pareto frontier.}
    \label{fig:ComparePareto}
\end{figure}

In Fig.~\ref{fig:ComparePareto}, we plot different NN hardware designs in the compute density (TOPS/mm$^2$) and energy efficiency (TOPS/W) space, including analog electronics~\cite{TrueNorth_2019, Neurogrid_2014, NN_ISCAS2010_Schemmel}, digital electronics~\cite{V100GPU, A100GPU, TPUV4, Graphcore_2002, Groq_2020}, and analog photonic tensor cores.
Note that PTCs are configured to have a single 64$\times$64 core, much smaller than electronics counterparts with multiple large-size (>1024) cores.
Analog electronics have relatively high energy efficiency with low compute density.
SoTA digital processors, e.g., TPUv4~\cite{TPUV4} and A100 GPU~\cite{A100GPU}, show comparable energy efficiency with around 1 TOPS/mm$^2$ area efficiency.
Analog photonic tensor cores outperform SoTA digital electronics by over two orders of magnitude in energy efficiency, while the compute density is still around 1 TOPS/mm$^2$.
With customized MOMMI devices, our \lemon designs, especially the \lemon(log) variant, shows 3-10 TOPS/mm$^2$ compute density, significantly advancing the Pareto frontier.
With more compact MMI designs and multiple wavelengths, the compute density of \lemon can potentially reach an even higher level.

\vspace{-15pt}
\subsection{System Throughput Comparison}
\vspace{-10pt}
We use an internal system-level photonic accelerator simulator to evaluate the throughput of different PTC designs in Fig.~\ref{fig:CompareFPS}.
The detailed architectural simulation is in Appendix~\ref{sec:Appendix:ArchSimulation}.
We adjust the core configurations to maintain similar area budgets for all PTCs for a fair comparison.
Our compact MOMMI-based design equipped with block unfolding allows more cores on chip while boosting the effective computing speed.
Our \lemon variants, on average, show 3.7-12$\times$ higher throughput (FPS) than baseline PTCs and 34.8-403$\times$ higher throughput than NVIDIA A100 GPU.

\begin{figure}
    \centering
    \includegraphics[width=0.95\columnwidth]{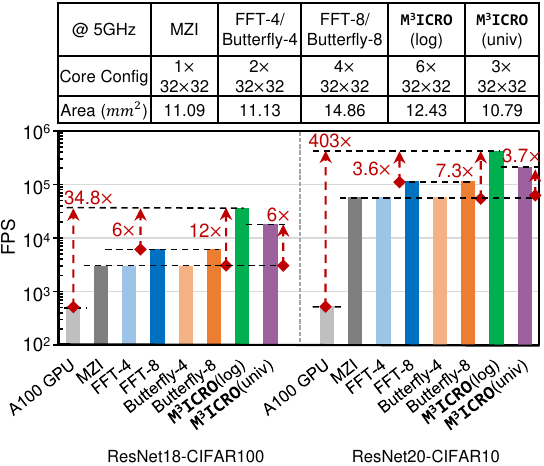}
    \caption{Compare the system-level throughput of the single-example inference task in frame-per-second (FPS) among PTCs (@5 GHz clock) and NVIDIA A100 GPU.
    The FPS of A100 is measured by the benchmarking tool from PyTorch with mixed precision.}
    \label{fig:CompareFPS}
\end{figure}

\vspace{-15pt}
\subsection{Discussion on Implementation of Programmable MOMMI}
\label{sec:DeviceImplementation}
\vspace{-10pt}
The practicality of index tuning for a multimode waveguide has been widely discussed in the literature~\cite{NP_OE2021_Larocque,NP_SPIE2019_Niekerk,NP_Micromachine2022_Chen}.
In this work, the MOMMI device is designed for weight-static linear transformation, which does not require high-speed modulation.
Hence we can use existing low-speed phase modulators as the tuning pads. 
For example, we can put thermal tuning pads on top of the multimode waveguide region, which has been experimentally demonstrated ~\cite{NP_SPIE2019_Niekerk,NP_Micromachine2022_Chen}.
To reduce the power consumption, we can also use non-volatile phase change materials (PCM)~\cite{NP_SciAdv2021_Delaney} or liquid crystal (LC)~\cite{NP_OE2021_Larocque} as the tuning pads with high index contrast and low static power consumption.
If the application requires high-speed weight reconfiguration, we can adopt electro-optic (EO) index-tuning materials, such as thin-film lithium niobate~\cite{NP_Nature2018_Wang}.

\vspace{-15pt}
\section{Conclusion}
\label{sec:Conclusion}
\vspace{-10pt}
In this study, we propose the first machine learning-enabled multi-path photonic tensor core \lemon based on customized programmable multi-operand multimode interference devices.
We thoroughly investigate its matrix expressivity and enable efficient PTC optimization with an ML-based training method.
We further introduce a block unfolding technique to enable full-range real-to-real linear transform for complex-valued PTC with 4 times higher efficiency than the differential photodetection approach.
Extensive evaluation shows that our customized \lemon PTC has close-to-digital task accuracy, 1.6-4.4$\times$ higher speed, 9.9-38.5$\times$ higher compute density, superior noise robustness, 3.7-12$\times$ higher system throughput than previous SoTA coherent PTCs, and 34.8-403$\times$ faster than A100 GPU.
This study opens up new possibilities for device customization and strengthens the integration of photonics and machine learning, driving the scalability and efficiency of photonic ML computing.

\vspace{-15pt}
\begin{acknowledgments}
\vspace{-10pt}
The authors acknowledge the Multidisciplinary University Research Initiative (MURI) program through the Air Force Office of Scientific Research (AFOSR), contract No.~FA 9550-17-1-0071, monitored by Dr. Gernot S. Pomrenke.
\end{acknowledgments}

\section{Data Availability}
The data that support the findings of this study are openly available in \href{https://github.com/JeremieMelo/M3ICRO-MOMMI}{M3ICRO-MOMMI}, reference number~\cite{MOMMI_Repo}.

\appendix
\begin{table}[]
\centering
\caption{Adopted component parameters in \lemon.
IL represents insertion loss.
}
\label{tab:device}
\resizebox{0.9\columnwidth}{!}{
\begin{tabular}{c|c|c}
\hline
Device                                    & Parameter   & Value                       \\ \hline
\multirow{2}{*}{Crossing (CR)~\cite{NP_OL2013_Zhang}}            & IL        & 0.02 dB                         \\
                                          & Area        & 7.4$\times$7.4 $\mu m^2$                      \\ 
                                           \hline
\multirow{4}{*}{Phase Shifter (PS)~\cite{NP_OE2021_Baghdadi}}            & IL        & 0.04 dB                         \\
                                          & Area        & 90$\times$40 $\mu m^2$                      \\ 
                                           & Response Time        & 10 $ns$                      \\ 
                                           & Static Power        & 0 $mW$                      \\ 
                                           \hline
\multirow{2}{*}{Y-branch (Y)~\cite{NP_IPJ2021_Nair}}                 & IL        & 0.3 dB                        \\
                                          & Area        & 1.8$\times$1.3 $\mu m^2$                     \\ \hline
\multirow{2}{*}{2$\times$2 Beam Splitter (BS)~\cite{NP_Thesis2007_Lin}}                 & IL        & 0.33 dB                        \\
                                          & Area        & 29.3$\times$2.4 $\mu m^2$                     \\ \hline
\multirow{2}{*}{4$\times$4 MMI~\cite{NP_Thesis2007_Lin}}                 & IL        & 0.33 dB                        \\
                                          & Area        & 55.4$\times$4.8 $\mu m^2$                     \\ 
                                          \hline
\multirow{3}{*}{MZM}                      & Tuning Power        & 2.25 mW~\cite{NP_OE2010_Dong}  \\
                                          & IL        & 1.2 dB~\cite{NP_OE2012_Suguru}                      \\
                                          & Area        & 260$\times$20 $\mu m^2$~\cite{NP_OE2012_Suguru} \\ \hline
\multirow{3}{*}{Photodetetcor~\cite{NP_optical2016_Huang}}             & Power & 1.1mW\\
                                          & Sensitivity & -25 dBm                      \\
                                          & Area & 4$\times$10 $\mu m^2$\\ \hline
\multirow{2}{*}{Laser~\cite{NP_CLEO2019_Wang}}            & Wall-plug efficiency        & 0.2                         \\
& Area        & 400$\times$300 $\mu m^2$                     \\\hline                                          
\end{tabular}
}
\vspace{-10pt}
\end{table}

\begin{table*}[]
\centering
\caption{Footprint ($A_{\text{core}}$), insertion loss ($IL_{\text{core}}$), and delay ($\tau_{\text{core}}$) analysis of photonic tensor cores.
$A$ is footprint, $IL$ is insertion loss, and $L$ is device length.
$W_0L_0$ is the area of a reference $k_0\times k_0$ MMI.
We assume the MMI size scales with $k^2$ based on Eq.~\eqref{eq:SelfImaging}.
FFT/Butterfly-$k'$ means that the PTC is of size-$k'$. If $k>k'$, the matrix is chunked into $(k/k')\times (k/k')$ blocks of size $k'\times k'$.
\#CR($k'$) and \#CCR($k'$) are the total crossing count and the number of cascaded crossings in the critical path.
$n_g$ is the group index and $c$ is the free-space light speed.
}
\label{tab:cost}
\resizebox{0.85\textwidth}{!}{
\begin{tabular}{c|c|c}
\hline
Design                                     & Metric & Value                       \\ \hline
\multirow{3}{*}{MZI~\cite{NP_NATURE2017_Shen}} & Footprint ($A_{\text{core}}$) &  $k^2(3A_{PS}+2A_{BS})$             \\
                     & IL ($IL_{\text{core}}$)  &  $(2k+1)(2IL_{BS}+2IL_{PS})$      \\
                     & Delay ($\tau_{\text{core}}$) &  $(2k+1)(2L_{BS}+2L_{PS})n_g/c$      \\\hline
FFT-$k'$~\cite{NP_ASPDAC2020_Gu} & Footprint ($A_{\text{core}}$)    &     $\ceil{k/k'}^2\big(k'(\log_2{k'}+2)A_{BS} + k'(2\log_2{k'}+2)A_{PS}+\#CR(k')\cdot A_{CR}\big)+2k(\ceil{k/k'}-1)A_{Y}$      \\
    Butterfly-$k'$~\cite{NP_ACSPhotonics2022_Feng}                 & IL ($IL_{\text{core}}$)   &  $2\ceil{\log_2(k/k')}IL_{Y}+(2\log_2{k'}+2)(IL_{BS}+IL_{PS})+\big(2\ceil{\log_2{(k/k')}}(k'-1)+\#CCR(k')\big)IL_{CR}$     \\
    & Delay ($\tau_{\text{core}}$)  &  $\big(2\ceil{\log_2(k/k')}L_{Y}+(2\log_2{k'}+2)(L_{BS}+L_{PS})+\big(2\ceil{\log_2{(k/k')}}(k'-1)+\#CCR(k')\big)L_{CR}\big)n_g/c$      \\\hline
\multirow{3}{*}{\lemon(log)} & Footprint ($A_{\text{core}}$)    &  $2\floor{\log_2{k}}L_{0}W_{0}k^2/k_0^2 + 4k(\floor{\log_2{k}}-1)(A_{PS}+A_{Y}) + 2kA_{Y}+k(k-1)A_{CR}$         \\
                     & IL ($IL_{\text{core}}$)   &   $2\cdot IL_{Y}+\floor{\log_2{k}}\cdot IL_{MMI}+(\floor{\log_2{k}}-1)(2IL_Y+IL_{PS})+2(k-1)IL_{CR}$    \\
                     & Delay ($\tau_{\text{core}}$)  &   $(2\cdot L_{Y}+\floor{\log_2{k}}\cdot L_{0}k/k_0+(\floor{\log_2{k}}-1)(2L_Y+L_{PS})+2(k-1)L_{CR})n_g/c$    \\\hline
\multirow{3}{*}{\lemon(univ)} & Footprint ($A_{\text{core}}$)   &  $PCL_{0}W_{0}k^2/k_0^2 + 2kP(C-1)(A_{PS}+A_{Y}) + 2(P-1)kA_{Y}+(P-1)k(k-1)A_{CR};~P,C=\eqref{eq:ParameterCount}$         \\
                     & IL ($IL_{\text{core}}$)  &  $2\ceil{\log_2{P}}\cdot IL_{Y}+C\cdot IL_{MMI}+(C-1)(2IL_{Y}+IL_{PS})+2\ceil{\log_2{P}}(k-1)IL_{CR};~P,C=\eqref{eq:ParameterCount}$     \\
                     & Delay ($\tau_{\text{core}}$)  &  $(2\ceil{\log_2{P}}\cdot L_{Y}+C\cdot L_{0}k/k_0+(C-1)(2L_{Y}+L_{PS})+2\ceil{\log_2{P}}(k-1)L_{CR})n_g/c;$     \\
                                          \hline
\end{tabular}
}
\end{table*}

\vspace{-15pt}
\section{Footprint, Insertion Loss, and Delay of Photonic Tensor Core}
\label{sec:Appendix:Footprint_Power}
\vspace{-10pt}
In Table~\ref{tab:cost}, we summarize the theoretical footprint $A_{\text{core}}$, insertion loss $IL_{\text{core}}$, and delay (latency) $\tau_{\text{core}}$ of different photonic tensor core designs with a core size of $k\times k$, which is used in Section~\ref{sec:HardwareEfficiency}.
The list of parameters used in the performance and efficiency calculation is in Table~\ref{tab:device}.
The $A_{\text{core}}$, $IL_{\text{core}}$, and $\tau_{\text{core}}$ are used in the calculation of total footprint, total insertion loss, and total delay in Section~\ref{sec:HardwareEfficiency}.

\vspace{-15pt}
\section{System-Level Performance Simulation}
\label{sec:Appendix:ArchSimulation}
\vspace{-10pt}
We adopt a system-level photonic accelerator simulator to simulate the performance and efficiency~\cite{NP_MLSysSNAP2023_Zhu}. 
The multi-core architecture has a DRAM, global SRAM buffer, input/activation SRAM buffers for each core, and multiple photonic tensor cores.
Optical interconnect is assumed for inter-core input operand broadcast.
The area, leakage power, and access energy of the 14 nm memory hierarchy are modeled by PCACTI~\cite{HWA_AVLSI2014_Shafaei}.
High-bandwidth memory (HBM) is used to supply data to the photonic system with a memory system bandwidth $>$1TB/s~\cite{DRAM}.
We use the ADC~\cite{NP_ISSCC2022_Liu} and DAC~\cite{NP_VLSI2020_Caragiulo} with 14 nm and 16 nm technology nodes, respectively, while their bit-widths and frequency are scaled accordingly~\cite{HWA_ISLPED2018_Kim}.
The scheduling of the multi-core accelerator adopts the weight-stationary dataflow to amortize the PTC weight reprogramming cost.

\section{Limitations and Future Directions}
Potential limitations of our proposed design are:
\begin{itemize}
    \item \textbf{Subspace tensor core}: our MOMMI's transfer matrix can not exactly express arbitrary matrix. 
    The relation between $\epsilon$ and transfer matrix $W$ is limited by the working principle of MMI.
    Theoretically, our multi-path PTC \lemon, though it can have multiple MMIs cascaded and connected in parallel, only numerically approximates a target matrix with a low error, which does not have a theoretical guarantee to express the exact matrix. Mapping a target matrix to our designs requires optimization-based mapping.
    \item \textbf{Weight-static tensor core}: the transfer matrix of M$^3$ICRO is not a simple explicit function of $\epsilon$, which is a complicated function learned by the neural network-based DPE. Hence, it can only be trained as a static weight for weight-stationary linear operation, e.g., $Z=Wx$ in linear/convolution layers in neural networks.
    A dynamic tensor product, e.g., $Z=XY$ in self-attention operations, cannot be realized, as the arbitrary dynamic tensor $X$ needs to be mapped to M$^3$ICRO through training, which cannot be efficiently performed in real time.
\end{itemize}
To further validate our design and improve its performance/efficiency, here are several future directions:
\begin{itemize}
    \item Experimentally demonstrate the usage of the proposed MOMMI and M$^3$ICRO PTC for real-world machine learning tasks.
    \item Explore other tuning mechanisms with different tuning pad geometries, locations, and materials, and customize the MMI structure to reduce static power consumption, reduce the size, and increase performance/robustness.
    \item Combine MOMMI with other multi-operand devices to realize more scalable tensor computations.
\end{itemize}

\section*{References}
\vspace{-20pt}
% \bibliography{./ref/Top_sim,./ref/Top,./ref/NN,./ref/NP}%

\newpage
\end{document}